# TIME AS AN ILLUSION

Paul S. Wesson

Department of Physics and Astronomy
University of Waterloo
Waterloo
Ontario  N2L 3G1
Canada

Revised Version: 25 April 2009

Abstract We review the idea, due to Einstein, Eddington, Hoyle and Ballard, that time is a subjective label, whose primary purpose is to order events, perhaps in a higher-dimensional universe.  In this approach, all moments in time exist simultaneously, but they are ordered to create the illusion of an unfolding experience by some physical mechanism.  This, in the language of relativity, may be connected to a hypersurface in a world that extends beyond spacetime.  Death in such a scenario may be merely a phase change.

1. <u>Introduction</u>

A couple of years after Einstein formulated special relativity, Minkowski in a famous speech argued that time should be welded to space to form spacetime. The result is a hybrid measure of separation, or interval, commonly called the Minkowski metric. It is the basis of quantum mechanics. By extension to curved as opposed to flat spacetime, we obtain a more complicated expression for the interval, which is the basis of cosmology. However, between the small systems of quantum theory and the large ones of cosmology, there are numerous others which can be adequately described by Newtonian mechanics and also involve time. An ongoing debate, in both philosophy and physics, has to do with the nature of time in its various applications. Especially: are the various usages of time in physics and everyday life consistent with a unique definition for it? Alternatively: while time occurs in many guises, what is the most useful way to view it at a conceptual level? We hope in what follows to answer these and related questions by re-examining the argument – espoused by Einstein, Eddington, Hoyle, Ballard and others – that time is essentially a subjective *ordering* device.

In doing this, it will be necessary to debunk certain myths about time, and to clarify statements that have been made about it. Certainly, time has been a puzzling concept throughout history. For example, Newton in his *Principia* (Scholium I), stated that "Absolute, true and mathematical time, of itself, and from its own nature, flows equably without relation to anything external, and by another name is called duration." This sentence is often quoted in the literature, and is widely regarded as being in opposition to the nature of time as embodied later in relativity. However, prior to that sentence, Newton also wrote about time and space that "… the common people conceive these quantities under no other notions but from the relation they bear to sensible objects." Thus Newton was aware that the "common" people in the 1700s held a view of time and other physical concepts which was essentially the same as the one used by Einstein, Minkowski, Poincaré and others in the 1900s as the basis for relativity.

As a property of relativity, it is unquestionably true that the time $t$ can be considered as a physical dimension, on the same basis as our measures $(x\,y\,z)$ of three-dimensional space. The result is spacetime. In this, the time part involves the product of $t$ with the speed of light $c$, which essentially transforms the 'distance' along the time axis to a length $ct$. Due to this, the interval is also a measure of which points are (or are not) in contact via the exchange of photons. Those particles with real interval can be in contact, while those with imaginary interval cannot be in contact.

This way of presenting Minkowski spacetime is conventional and familiar. However, it has a corollary which is not so familiar: particles with zero interval are *coincident* in 4D. Einstein realized this, and it is the basis of his definition of simultaneity. But it is not a situation which most people find easy to picture, so they decompose 4D spacetime into 3D space and 1D time, and visualize a photon propagating through $x\,y\,z$ over time $t$. Eddington, the noted contemporary of Einstein, also appreciated the subjective nature of the situation just described, and went on to argue that much of what is called objective in physics is in fact subjective or invented. The speed of light was also commented on later by a few deep thinkers such as McCrea and Hoyle, who regarded it as a mere man-made constant. From the Eddington viewpoint, one can argue that the decomposition of 4D

Minkowski spacetime into separate 3D and 1D parts is a subjective act, so that in effect the photon has been invented as a consequence of separating space and time.

Below, we will enlarge on the possibly subjective nature of physics, with an emphasis on the concept of time. We will in fact suggest that time is a subjective ordering device, used by humans to make sense of their world. Several workers have expressed this idea, including Einstein (1955 in Hoffmann 1972), Eddington (1928, 1939), Hoyle (1963, 1966), Ballard (1984) and Wesson (2001). We hope to show that this approach makes scientific sense, and from a common-day perspective has certain comforts.

Such an approach is, however, somewhat radical. So to motivate it, we wish to give a critique of other, more mainstream views. This will be short, because good reviews of the nature of time are available by many workers including Gold (1967), Davies (1974), Whitrow (1980), McCrea (1986), Hawking (1988), Landsberg (1989), Zeh (1992), Woodward (1995) and Halpern and Wesson (2006). We will discuss contending views of the nature of time in Section 2, introduce what seems to be a better approach in Section 3, and expand on the implications of this in Section 4. Although it is not essential, it will become apparent that our new approach to time is psychologically most productive when the world is taken to have more dimensions than the four of spacetime, in accordance with modern physics.

2.  Physics and the Flow of Time

The idea that time flows from the past to the future, and that the reason for this has something to do with the natural world, has become endemic to philosophy and phys-ics. However, this idea is suspect. We will in this section examine briefly the three ways in which the direction of time's 'arrow' is commonly connected with physical processes, and argue that they are all deficient. Quite apart from technical arguments, a little thought will show that a statement such as the "flow of time", despite being everyday usage, is close to nonsensical. For the phrase implies that time itself can be measured with respect to another quantity of the same kind. This might be given some rational basis in a multidimensional universe in which there is more than one time axis (see below); but the everyday usage implies measuring the change of a temporal quantity against itself, which is clearly a contradiction in terms. Such a sloppy use of words appears to be tolerated because there is a widespread belief that the subjective, unidirectional nature of time can be justified by more concrete, physical phenomena.

Entropy is a physical concept which figures in the laws of thermodynamics. Strictly speaking, it is a measure of the number of possible states of a physical system. But more specifically, it is a measure of the disorder in a system; and since disorder is observed to increase in most systems as they evolve, the growth of entropy is commonly taken as indicative of the passage of time. This connection was made by Eddington, who also commented on the inverse relationship between information and entropy (Eddington 1928, 1939). However, the connection has been carried to an unreasonable degree by some subsequent writers, who appear to believe that the passage of time is equivalent to the increase of entropy. That this is not so can be seen by a simple counter-argument: If it were true, each person could carry a badge that registered their entropy, and its measurement would correlate with the time on a local clock. This is clearly daft.

A more acceptable application of the notion of entropy might be found in the many-worlds interpretation of quantum mechanics. This was proposed by Everett (1957), and supported as physically reasonable by De Witt (1970). In it, microscopic systems bifurcate, and so define the direction of the future. In principle, this approach is viable. However, the theory would be better couched in terms of a universe with more than the four dimensions of spacetime; and interest in the idea of many worlds appears to have lapsed, because there is no known way to validate or disprove their existence.

Another physical basis for the passage of time which has been much discussed concerns the use of so-called retarded potentials in electromagnetism. The connection is somewhat indirect, but can be illustrated by a simple case where light propagates from one point to another. (This is what happens when humans apprehend things by the sense of sight, and is also how most information is transmitted by modern technology.) Let the signal be emitted at point P and observed at point O, where the distance between them is $d$ and is traversed at lightspeed $c$. Now Maxwell's equations, which govern the interaction, are symmetric in the time $t$. (We are assuming that the distance is small enough that ordinary three-dimensional space can be taken as Euclidean or flat.) However, in order to get the physics right, we have to use the electromagnetic potential not at time $t$ but at the retarded time $(t\text{-}d/c)$. This is, of course, the time 'corrected' for the travel lag from the point P of emission to the point O of observation. Such a procedure may appear logical; but it has been pointed out by many thinkers that it automatically introduces a time asymmetry into the problem (see Davies 1974 for an extensive review). The use of retarded potentials, while they agree with observations, is made even more puzzling by the fact that Maxwell's equations are equally valid if use is made instead of the 'advanced' potentials defined at $(t + d/c)$. In short, the underlying theory treats negative and positive increments of time on the same footing, but the real world appears to prefer the solutions where the past evolves to the future. Studies have been made of the symmetric case, called Wheeler/Feynman electrodynamics, where both retarded and advanced potentials are allowed. One argument for why we do not experience the signals corresponding to the advanced potentials is that due to Hoyle and Narlikar (1974). They reasoned that the advanced signals would be absorbed in certain types of cosmological model, leaving us with a universe which is apparently asymmetric between the past and future. This explanation is controversial, insofar as it appeals to unverified aspects of the large-scale cosmos. On the small scale, it appears that the need for retarded potentials in electrodynamics leads to a locally-defined arrow of time; though whether this is due to objective physical reasons, or to some subjective bias on our part, remains obscure.

The big bang offers yet another way of accounting for the arrow of time. According to Einstein's theory of general relativity, everything we observe came into existence in a singularity at a specific epoch, which supernova data fix at approximately $13 \times 10^9$ years before the present. This description is familiar to all, and carries with it the implication that the universe in a dynamical sense has a preferred direction of evolution. However, closer examination shows that it is really the recession of the galaxies from each other, rather than the big bang, which identifies the time-sense of the universe's evolution. This was understood by Bondi (1952), who was one of the founders with Gold and Hoyle of the steady-state theory. In it, matter is continuously created and condenses to form new galaxies, whose average density is thereby maintained even though the whole system is expanding. While no longer regarded as a practical cosmology, the steady-state theory shows that it is the motions of galaxies which essentially defines a preferred direction for time, rather than the (still poorly understood) processes by which they may have formed after the big bang. Let us, in fact, temporarily forget about the latter event, and consider an ensemble of gravitating galaxies. Then there are in principle only three modes of evolution: expansion, contraction and being static. The last can be ruled out, because it is widely acknowledged that such a state, even if it existed, would be unstable and tip into one of the other two modes. We are thus lead to the realization that if the arrow of time is dictated by the dynamical evolution of the universe, its sense is given *a priori* by a 50/50 choice analogous to flipping a cosmic coin. That is, there is no dynamical reason for believing that events should go forward rather than backwards in time. In addition to this, there is also the problem that there is no known physical process which can transfer a cosmic effect on a lengthscale of $10^{18}$ cm down to a human one of order $10^2$ cm. In order to circumvent this objection, it has been suggested that the humanly-perceived arrow of time is connected instead to smaller-scale astrophysics, such as the nucleosynthesis of elements that determines the evolution of the Sun. This process might, via the notion of entropy as discussed above, be connected to geophysical effects on the Earth, and so to the biology of its human inhabitants. But it is really obvious, when we pick apart the argument, that there is no discernable link between the mechanics of the evolving universe and the sense of the passage of time which is experienced by people.

The preceding issues, to do with entropy, electrodynamics and cosmology, have the unfortunate smell of speculation. Dispassionate thought reveals little convincing connection between the time coordinate used in physics and the concept of age as used in human biology. We can certainly *imagine* possible connections between physical and human time, as for example in *Einstein's Dreams* by Lightman (1993). There, the effects of relativity such as time dilation are described in sociological contexts. But, there is a large gap between the fluid manner in which time can be manipulated by the novelist and the rigid transformations of time permitted to the physicist. Indeed, while the physicist may be able to handle the "$t$" symbol in his equations with dexterity, he looks clumsy and strained when he attempts to extend his theories to the practicality of everyday existence. That is why the sayings about time by physicists mainly languish in obscurity, while those by philosophers have wider usage.

In the latter category, we can consider the statement of Marcel Proust: "The world was not created at the beginning of time. The world is created every day." This appears to dismiss the big bang, and by implication other parts of physics, as irrelevant to the human experience of time. However, it is more rewarding to consider statements like the foregoing as pointed challenges to the physicist. To be specific: Is there a view of "time" which is compatible with the rather narrow usage of the word in physics, and yet in agreement with the many ways in which the concept is experienced by people?

3. <u>Time as a Subjective Ordering Device</u>

The differing roles which time plays in physics and everyday life has led some workers to the conclusion that it is a subjective concept. Let us consider the following quotes:

Einstein (as reported by Hoffman): "For us believing physicists the distinction between past, present and future is only an illusion, even if a stubborn one."

Eddington: "General scientific considerations, favour the view that our feeling of the going on of time is a sensory impression; that is to say, it is as closely connected with stimuli from the physical world as the sensation of light is. Just as certain physical disturbances entering the brain cells via the optic nerves occasion the sensation of light, so a change of entropy … occasions the sensation of time succession, the moment of greater entropy being *felt* to be the later."

Hoyle: "All moments of time exist together." "There is no such thing as 'waiting' for the future." "It could be that when we make subjective judgments we're using connections that are non-local … there is a division, the world divides into two, into two completely disparate stacks of pigeon holes."

Ballard: "Think of the world as a simultaneous structure. Everything that's ever happened, all the events that *will* ever happen, are taking place together." "It's possible to imagine that everything is happening at once, all the events 'past' and 'future' which constitute the universe are taking place together. Perhaps our sense of time is a primitive mental structure that we inherited from our less intelligent forbears."

The preceding four opinions about time have an uncanny similarity, given that they apparently originate independently of each other. However, they are all compatible with Eddington's view of science, wherein certain concepts of physics are not so much discovered as invented (see Wesson 2000 for a short review). The subjective nature of time is also compatible with current views of particle physics and cosmology, wherein several worlds exist alongside each other (Everett 1957, De Witt 1970, Penrose 1989, Wesson 2006, Petkov 2007). It is important to realize that there need not be anything mystical about this approach. For example, Hoyle considers a 4D world of the usual type with time and space coordinates $t$ and $x\,y\,z$ which define a surface $\phi(t, x\,y\,z) = C$. Here $C$ is a parameter which defines a subset of points in the world. Changing $C$ changes the subset, and "We could be said to live our lives through changes of $C$." In other words, the life of a person can be regarded as the consequence of some mechanism which picks out sets of events for him to experience.

What such a mechanism might be is obscure. Hoyle speculated that the mechanism might involve known physical fields such as electromagnetism, which is the basis of human brain functions. It might plausibly involve quantum phenomena, amplified to macroscopic levels by the brain in the manner envisaged by Penrose (1989). However, while the precise mechanism is unknown, some progress can be made in a general way by noting that Hoyle's $C$-equation above is an example of what in relativity is known as a *hypersurface*. This is the relation one obtains when one cuts through a higher-dimensional manifold, defining thereby the usual 4D world we know as spacetime. It is in fact quite feasible that the Minkowski spacetime of our local experience is just a slice through a world of more than 4 dimensions.

In fact, higher dimensions are the currently popular way to unify gravity with the interactions of particle physics, and reviews of the subject are readily available (e.g., Wesson 2006 from the physical side and Petkov 2007 from the philosophical side). Since we are here mainly interested in the concept of time, let us concentrate on one exact solution of the theory for the simplest case when there is only a single extra dimension. (See Wesson 1999 for a compendium of higher-dimensional solutions including the one examined here.) Let us augment the time (*t*) and the coordinates of Euclidean space (*x y z*) by an extra length (*l*). Then by solving the analog of Einstein's equations of general relativity in 5D, the interval between two nearby points can be written

$$dS^2 = l^2 dt^2 - l^2 \exp i(\omega t + k_x x) dx^2 - l^2 \exp i(\omega t + k_y y) dy^2 - l^2 \exp i(\omega t + k_z z) dz^2 + L^2 dl^2. \quad (1)$$

Here $\omega$ is a frequency, $k_x$ *etc.* are wave numbers and *L* measures the size of the extra dimension. This equation, while it may look complicated, has some very informative aspects: (*a*) it describes a wave, in which parts of what are commonly called space can come into and go out of existence; (*b*) it can be transformed by a change of coordinates to a flat manifold, so what looks like a space with structure is equivalent to one that is featureless; (*c*) the signature is + – – – +, so the extra coordinate acts like a second time. These properties allow of some inferences relevant to the present discussion: (*a*) even ordinary 3D space can be ephemeral; (*b*) a space may have structure which is not intrinsic but a result of how it is described; (*c*) there is no unique way to identify time.

This last property is striking. It means that in grand-unified theories for the forces of physics, the definition of time may be ambiguous. This classical result confirms the inference from quantum theory, where the statistical interaction of particles can lead to thermodynamic arrows of time for different parts of the universe which are different or even opposed (Shulman 1997, 2000). It should be noted that the existence of more than one 'time' is not confined to 5D relativity, but also occurs in other $N$-dimensional accounts such as string theory (Bars et al. 1999). Indeed, there can in principle be many time-like coordinates in an $N$-dimensional metric.

In addition, the definition of time may be altered even in the standard 4D version of general relativity by a coordinate transformation. (This in quantum field theory is frequently called a gauge choice.) The reason is that Einstein's field equations are set up in terms of tensors, in order to ensure their applicability to any system of coordinates. This property, called covariance, is widely regarded as essential for any modern theory of physics. However, if we wish to have equations which are valid irrespective of how we choose the coordinates, then we perforce have to accept the fact that time and space are malleable. Indeed, covariance even allows us to *mix* the time and space labels. Given the principle of covariance, it is not hard to see why physicists have abandoned the unique time label of Newton, and replaced it by the ambiguous one of Einstein.

We are led to the realization that the concept of time is as much a puzzle to the physicist as it is to the philosopher. Paradoxically, the average person in the street probably feels more comfortable about the issue than those who attempt to analyse it.

However, it is plausible that time in its different guises is a device used by people to organize their existence, and as such is at least partially subjective in character.

4.  Mathematics and Reality

In the foregoing, we saw that several deep thinkers have arrived independently at a somewhat intriguing view of time. To paraphrase them: time is a stubborn illusion (Einstein), connected with human sensory impressions (Eddington), so that all moments of time exist together (Hoyle), with the division between past and future merely a holdover from our primitive ancestors (Ballard). Perhaps the most trenchant opinion is that of Hoyle (1966), who summarizes the situation thus: "There's one thing quite certain in this business. The idea of time as a steady progression from past to future is wrong. I know very well we feel this way about it subjectively. But we're all victims of a confidence trick. If there's one thing we can be sure about in physics, it is that all times exist with equal reality"

This view of time can be put on a physical basis. We imagine that each person's experiences are a subset of points in spacetime, defined technically by a hypersurface in a higher-dimensional world, and that a person's life is represented by the evolution of this hypersurface. This is admittedly difficult to visualize. But we can think of existence as a vast ocean whose parts are all connected, but across which a wave runs, its breaking crest precipitating our experiences.

A mathematical model for a wave in five dimensions was actually considered in the preceding section as Equation (1). It should be noted that there is nothing very spe-cial about the dimensionality, and that it is unclear how many dimensions are required to adequately explain all of known physics. The important thing is that if we set the interval to zero, to define a world whose parts are connected in higher dimensions, then we necessarily obtain the hypersurface which defines experience in the lower-dimensional world. It is interesting to note that the behaviour of that hypersurface depends critically on the number of plus and minus signs in the metric (i.e., on the signature). In the canonical extension of Einstein's theory of general relativity from four to five dimensions, the hypersurface has two possible behaviours. Let us express the hypersurface generally as a length, which depends on the interval of spacetime $s$, or equivalently on what physicists call the proper time (which is the time of everyday existence corrected to account for things like the motion). Then the two possible behaviours for the hypersurface may be written

$$l = l_o \exp(s/L) \quad \text{and} \quad l = l_o \exp(is/L) \quad . \tag{2}$$

Here $l_o$ is a fiducial value of the extra coordinate, $L$ is the length which defines the size of the fifth dimension, and $s$ is the aforementioned interval or proper time. The two noted behaviours describe, respectively, a growing mode and an oscillating mode. The difference between the two modes depends on the signature of the metric, and is indicated by the absence or presence of $i \equiv \sqrt{-1}$ in the usual manner. So far, the analysis follows the basic idea about experience due to Hoyle but expressed in the language of hypersurfaces as discussed by Wesson (see Hoyle and Hoyle 1963, Wesson 2006). However, it is possible to go further, and extend the analysis into the metaphysical domain for those so inclined. This by virtue of a change from the growing mode to the oscillatory mode, with the identification of the former with a person's material life and the latter with a person's spiritual life. That is, we obtain a simple model wherein existence is described by a hypersurface in a higher-dimensional world, with two modes of which one is growing and is identified with corporeal life, and one is wave-like and is identified with the soul, the two modes separated by an event which is commonly called death.

Whether one believes in a model like this which straddles physics and spirituality is up to the individual. (In this regard, the author is steadfastly neutral.) However, it is remarkable that such a model can even be formulated, bridging as it does realms of experience which have traditionally been viewed as immutably separate. Even if one stops part way through the above analysis, it is clear that the concept of time may well be an illusion. This in itself should be sufficient to comfort those who fear death, which should rather be viewed as a phase change than an endpoint.

5.  Conclusion

Time is an exceptionally puzzling thing, because people experience it in different ways. It can be formalized, using the speed of light, as a coordinate on par with the coordinates of ordinary three-dimensional space. But while spacetime is an effective tool for the physicist, this treatment of time seems sterile to the average person, and does not explain the origin of time as a concept. There are shortcomings in purely physical explanations of time and its apparent flow, be they from entropy, many-worlds, electromagnetism or the big bang. Such things seem too abstract and remote to adequately explain the individual's everyday experience of time. Hence the suggestion that time is a subjective ordering device, invented by the human mind to make sense of its perceived world.

This idea, while not mainstream, has occurred to several thinkers. These include the philosopher Proust, the scientists Einstein, Eddington and Hoyle, and the novelist Ballard. It is noteworthy that the idea appears to have its genesis independently with these people. And while basically psychological in nature, it is compatible with certain approaches in physics, notably Penrose's suggestion that the human brain may be a kind of amplification organ for turning tiny, quantum-mechanical effects into measurable, macroscopic ones. The idea of time as an ordering device was given a basis in the physics of relativity by Hoyle, who however only sketched the issue, arguing that the movement of a hypersurface would effectively provide a model for the progress of a person's life. This approach can be considerably developed, as outlined above, if we assume that the experience-interface is related to a 4D hypersurface in a 5(or higher)D world. Then it is possible to write down an equation for the hypersurface, which can have an evolutionary and an oscillatory phase, which might (if a person is so inclined) be identified with the materialistic and spiritual modes of existence. Perhaps more importantly, in this 5D approach, the interval (or 'separation') between points is zero, so all of the events in the world are in (5D) causal contact. In other words, everything is occurring simultaneously.

That this picture may be difficult to visualize just bolsters the need for something like the concept of time, which can organize the simultaneous sense data into a comprehensible order.

Time, viewed in this manner, is akin to the three measures of ordinary space, at least insofar as how the brain works. Humans have binocular vision, which enables them to judge distances. This is an evolutionary, biological trait. Certain other hunting animals, like wolves, share it. By comparison, a rabbit has eyes set into the sides of its head, so while it can react well to an image that might pose a threat, it cannot judge distance well. But even a human with good vision finds it increasingly difficult to judge the relative positions of objects at great distance: the world takes on a two-dimensional appearance, like a photograph, or a landscape painting. In the latter, a good artist will use differing degrees of shade and detail to give an impression of distance, as for example when depicting a series of hills and valleys which recede to the horizon. Likewise, the human brain uses subtle clues to do with illumination and resolution to form an opinion about the relative spacing of objects at a distance. This process is learned, and not perfectly understood by physiologists and psychologists; but is of course essential to the adequate functioning of an adult person in his or her environment. Astronomers have long been aware of the pitfalls of trying to assess the distances of remote objects. Traditionally, they measured offsets in longitude and latitude by means of two angles indicated by the telescope, called right ascension and declination. But they had no way of directly measuring the distances along the line of sight, and so referred to their essentially 2D maps as being drawn on the surface of an imaginary surface called the celestial sphere. Given such a limited way of mapping, it was very hard to decide if two galaxies seen close together on the sky were physically close or by chance juxtaposed along the line of sight. In lieu of a direct method of distance determination, astronomers fell back on probability arguments to decide (say) if two galaxies near to each other on a photographic plate were really tied together by gravity, or merely the result of a coincidental proximity in 2D while being widely separated in 3D. The situation changed drastically when technological advances made it easier to measure the redshifts of galaxies, since the redshift of a source could be connected via Hubble's law to the physical distance along the line of sight. Thus today, combining angular measurements for longitude and latitude with redshifts for outward distance, astronomers have fairly good 3D maps of the distribution of galaxies in deep space.

In effect, astronomers have managed to replace the photograph (which is essentially 2D) by the hologram (which provides information in 3D). However, whether this is done for a cluster of galaxies or a family portrait, the process of evaluating distance is a relatively complicated one. The human brain evaluates 3D separations routinely, and we are not usually aware of any conscious effort in doing so. But this apparently mundane process is also a complicated one. If we take it that the concept of time is similar to the concept of space, it is hardly surprising that the human brain has evolved its own subtle way of handling 'separations' along the time axis of existence.

Thus the idea of time as a kind of subjective ordering device, by which we make sense of a simultaneous world, appears quite natural.